\documentclass[manuscript,screen]{acmart}
\AtBeginDocument{%
  \providecommand\BibTeX{{%
    \normalfont B\kern-0.5em{\scshape i\kern-0.25em b}\kern-0.8em\TeX}}}

\copyrightyear{2024}
\acmYear{2024}
\setcopyright{acmlicensed}\acmConference[CHI '24]{Proceedings of the CHI Conference on Human Factors in Computing Systems}{May 11--16, 2024}{Honolulu, HI, USA}
\acmBooktitle{Proceedings of the CHI Conference on Human Factors in Computing Systems (CHI '24), May 11--16, 2024, Honolulu, HI, USA}
\acmDOI{10.1145/3613904.3642224}
\acmISBN{979-8-4007-0330-0/24/05}

\acmSubmissionID{5510}

\usepackage{xspace}
\usepackage[normalem]{ulem}
\usepackage{tabularx}
\usepackage{graphicx}
\usepackage{booktabs}
\begin{document}


\title{\tool: Supporting Creative Interior Color Design Ideation with Large Language Model}

\author{Yihan Hou}
\affiliation{%
  \institution{The Hong Kong University of Science and Technology (Guangzhou)}
  \country{China}
}
\author{Manling Yang}
\affiliation{%
  \institution{The Hong Kong University of Science and Technology (Guangzhou)}
  \country{China}
}
\author{Hao Cui}
\affiliation{%
  \institution{The Hong Kong University of Science and Technology (Guangzhou)}
  \country{China}
}
\author{Lei Wang}
\affiliation{%
  \institution{China Academy of Art}
  \country{China}
}
\author{Jie Xu}
\affiliation{%
  \institution{China Academy of Art}
  \country{China}
}
\author{Wei Zeng}
\authornote{Wei Zeng is the corresponding author.}
\affiliation{%
  \institution{The Hong Kong University of Science and Technology (Guangzhou)}
  \country{China}
}
\affiliation{%
  \institution{The Hong Kong University of Science and Technology}
  \country{Hong Kong SAR, China}
}

\newcommand{\outline}[1]{\colorbox{mygreen}{\textcolor{white}{#1}}}
\newcommand{\llmop}[1]{\textcolor{llmop_color}{\textbf{#1}}}
\newcommand{\interaction}[1]{\textcolor{userinter_color}{\textbf{#1}}}
\newcommand{\re}[1]{\textcolor{revision}{#1}}

\definecolor{mygreen}{RGB}{128,164,146}
\definecolor{myred}{RGB}{243, 166, 148}
\definecolor{llmop_color}{RGB}{128, 164, 146}
\definecolor{userinter_color}{RGB}{122, 123, 120}
\definecolor{revision}{RGB}{0,0,0}

\newcommand{\tool}{\emph{C2Ideas}\xspace}

\newcommand{\q}[1]{\textit{``#1''}}
\newcommand{\qn}[1]{``#1''}
\newcommand{\eg}{\emph{e.g.}}
\newcommand{\ie}{\emph{i.e.}}

\newcommand{\strike}[1]{\textcolor{red}{\sout{#1}}}
\newcommand{\strikeg}[1]{\textcolor{blue}{\sout{#1}}}
\newcommand{\add}[1]{\textcolor{red}{#1}}
\newcommand{\replace}[2]{\strikeg{#1 }\add{#2}}

\begin{abstract}
Interior color design is a creative process that endeavors to allocate colors to furniture and other elements within an interior space.
While much research focuses on generating realistic interior designs, these automated approaches often misalign with user intention and disregard design rationales.
Informed by a need-finding preliminary study, we develop \tool, an innovative system for designers to creatively ideate color schemes enabled by an intent-aligned and domain-oriented large language model.
\tool integrates a three-stage process: \emph{Idea Prompting} stage distills user intentions into color linguistic prompts; \emph{Word-Color Association} stage transforms the prompts into semantically and stylistically coherent color schemes; and \emph{Interior Coloring} stage assigns colors to interior elements complying with design principles. 
We also develop an interactive interface that enables flexible user refinement and interpretable reasoning. 
\tool has undergone a series of indoor cases and user studies, demonstrating its effectiveness and high recognition of interactive functionality by designers.
\end{abstract}

\ccsdesc[500]{Human-centered computing~User interface toolkits}

\keywords{large language model, color deisgn}




\maketitle

\section{Introduction}
\label{sec:introduction}
Design ideation stands as a foundational pillar in the realm of creative activities~\cite{koch2019may}.
Recent advances in artificial intelligence (AI) technologies have opened up unprecedented avenues for the enhancement of creative design processes~\cite{shi2023understanding}.
Interior color design, with the goal of creating an aesthetically pleasing and emotionally impactful environment, is a typical creative design task benefiting from the development of AI~\cite{gbr2023robotecture}.
\re{
    Incorporating AI into professional interior design can provide significant benefits for designers.
    Utilizing data-driven insights, AI can help designers align their decisions more closely with the client's abstract requirements, thereby reducing the impact of personal bias~\cite{matthiesen2021approach, lee2022conceptual}.
    In addition, AI can streamline the design process, especially in the initial stages of a designer's search for inspiration,
    thus reducing the time and effort usually spent sifting through a variety of irrelevant sources.}
Nevertheless, existing AI-based approaches for color design have been focusing on utilizing pre-trained models learned from large datasets \cite{solah2022mood, chen2023generating}, whilst few efforts are devoted to understanding such abstract requirements.

\begin{figure}[t]
    \centering
    \includegraphics[width=0.99\linewidth]{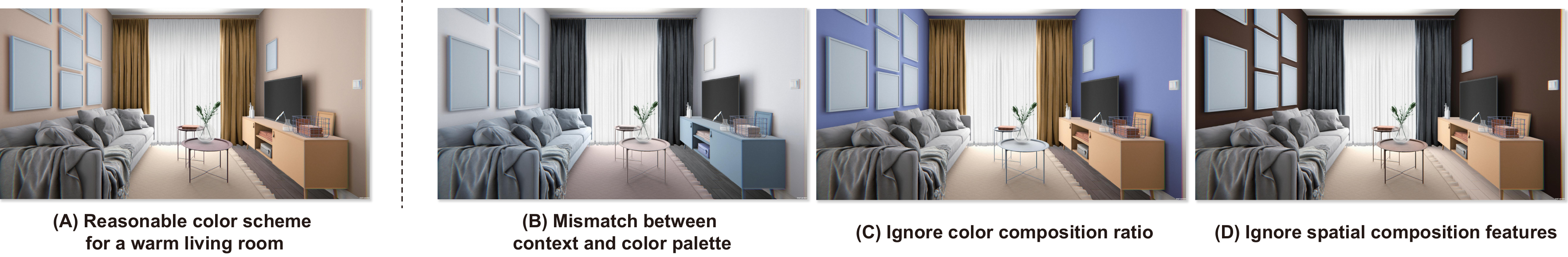}
    \vspace{-1mm}
    \caption{Examples of successful interior color design (A) aligned with the intention of creating a \q{warm living room}, and
        common pitfalls of contextually misaligned color scheme (B), improper color composition ratios (C), and no coherence with the interior elements (D).}
    \vspace{-2mm}
    \label{fig:pitfalls}
\end{figure}

There is a high demand for an AI-assisted tool that offers flexibility and control \re{to assist designers} in interior color design.
Such a tool should be controllable to meet designers' needs while also adhering to the design rationales of interior color design.
Nevertheless, developing such a tool comes with several challenges: 1) clients often communicate their requirements ambiguously, making interpretation complex~~\cite{xu2014voyant}, and 2) creating a coherent interior color design is challenging due to the intricate interplay of color compositions and furniture layout.
Failing to address these challenges can often lead to design pitfalls, as illustrated in Figure~\ref{fig:pitfalls}.
For instance, when clients express a desire for a \q{warm living room}, a misunderstanding of this requirement can result in a color scheme that is contextually mismatched (Figure~\ref{fig:pitfalls} (B)).
Even with correct color schemes, improper ratios in color composition (Figure~\ref{fig:pitfalls} (C)) or neglecting coherence with the interior elements (Figure~\ref{fig:pitfalls} (D)), can create uncomfortable perceptions of the design.

Numerous efforts have been made to intertwine AI and color design, encompassing approaches that range from data-driven models learning textual-color mappings~\cite{solah2022mood, chen2016stylistic} to prompt-driven strategies guiding pre-trained models towards user-desired outcomes~\cite{chen2023generating}.
However, these methods encounter challenges when handling text inputs, which are the primary means for humans to express their design intentions, particularly during the ideation stage.
Moreover, these methodologies typically employ an end-to-end approach, directly translating user input into color schemes~\cite{chen2016stylistic, solah2022mood}, often resulting in designs that are challenging for professionals to modify~\cite{chen2023generating}.
The emergence of large language models (LLMs) offers new avenues for creative design due to their ease of use and powerful generation capabilities \cite{stevenson2022putting}.
Their potential extends to comprehending user requirements and providing aligned feedback.
However, though some recent works have adopted LLMs in creative design fields, such as game design~\cite{lanzi2023chatgpt} and chart coloring~\cite{shi2023nl2color}, the application to interior color design remains largely unexplored and challenging.
First, LLM is inherently an end-to-end approach, producing homogeneous content for identical inputs, which may not be conducive to design ideation and lacks support for addressing the controllability issues in the design process.
Second, being a generic model, LLM lacks domain-specific knowledge, such as color psychology, the natural color system (NCS), and composition principles, all of which are crucial aspects of interior color design.
Existing research often falls short in terms of comprehensive integration of domain knowledge.

To address these challenges, we introduce \tool, an LLM-based color design system with twofold considerations:
1) \emph{Rationality}: Integrating diverse domain-specific knowledge in the color design process, and 2) \emph{Controllability}: Granting users flexibility in expressing their design intent throughout the process.
The design of \tool draws insights from a formative study that delves into comprehending the domain knowledge and design process of interior color design, along with expert perspectives on integrating AI assistance into the design process (Sect.~\ref{sec:preliminary}).
Then, we leverage the idea of the chain-of-thought (CoT)~\cite{wei2022chain}, to decompose the color design process into consecutive steps and apply the domain knowledge to each step (Sects.~\ref{ssec:prompting} $-$~\ref{ssec:interior_coloring}).
Multi-level interactions that allow users to intervene in the design process are incorporated at each step to support system controllability.
We also develop an interactive interface for user refinement, enabling a semi-automated method for converting user intent into scene-specific color schemes (Sect.~\ref{ssec:system_design}).
We evaluate the effectiveness of our approach through a user study, and the results showed that our method outperforms a baseline LLM method and other alternatives that do not involve domain knowledge.
We also conduct an expert interview to evaluate the controllability and usability of our system, and the results showed that our system is easy to use and the system can meet the different needs of designers to express their intentions (Sect.~\ref{sec:evaluation}).

In summary, our work makes the following contributions:
\begin{enumerate}
    \item We identify and distill design considerations for LLM-based assistants in interior color design, emphasizing design rationality and system controllability.

    \item \re{We introduce an LLM-based system workflow for interior color design, which incorporates domain knowledge through prompt engineering and effectively aligns with diverse user intents by employing a multi-tiered feedback mechanism.}

    \item We develop a semi-automated approach \tool, that converts user intent into scene-specific color schemes bolstered by an interactive interface for user refinement.
          The effectiveness and usability of \tool are affirmed through user studies and expert interviews.
\end{enumerate}
\section{Related Works}
\subsection{Creative Color Design}
Creative color design entails a deliberate and strategic choice of colors to fulfill specific aesthetic and functional requirements, often culminating in customized color schemes tailored to a particular context or set of criteria~\cite{al2016influence, zhang2020research, li2016creative}.
As color is a foundational visual channel, creative color design transcends various fields, including visualization~\cite{shi2022colorcook, shi2023nl2color}, graphical design~\cite{qiu2022intelligent,shi2023stijl}, and interior design~\cite{solah2022mood}.
To craft coherent and visually pleasing color schemes, designers draw upon principles such as color harmony~\cite{shen2000color, lin2020c3}, color psychology~\cite{ou2004study, kurt2014effects}, and color compatibility~\cite{chen2016stylistic}.
Building upon these principles, computer-aided color design has emerged, utilizing rule-based approaches that encapsulate design considerations like harmony~\cite{lin2020c3} and color compatibility~\cite{chen2016stylistic} into algorithmic rules.
Meeting these optimization criteria can be challenging, as the parameters for color design vary upon factors such as application domains and personalized preferences~\cite{shi2022colorcook}.
For instance, individuals may have diverse preferences when it comes to decoration styles~\cite{chen2016stylistic}, and these preferences can vary significantly between interior and graphic design contexts.

Interior color design, as a subfield, presents unique challenges due to a multitude of factors.
In practice, interior color design aims to cater to a broad range of users, with their intentions expressed in vague and ambiguous textual descriptions.
Traditional rule-based color design methods often struggle to interpret these textual descriptions effectively.
Studies have turned to crowd-sourced experiments~\cite{mohammad2013colourful, kim2020lexichrome,shi2022colorcook} and image-text corpora~\cite{setlur2015linguistic,bahng2018coloring,ikoma2020effect} to better understand the intricate relationship between color perceptions and textual descriptions.
However, these approaches often have limitations in terms of the number of text-color pairs, with some only supporting texts containing specific entities (\eg,~\cite{setlur2015linguistic}) or a limited range of categorical colors (\eg,~\cite{kim2020lexichrome}).
Recent advances in text-to-image generative models such as Stable Diffusion~\cite{rombach2022high} and DALL E2~\cite{ramesh2022hierarchical} have addressed data cost and style inflexibility issues.
Nevertheless, the generative models may produce inappropriate space dimensions and unrealistic outputs~\cite{chen2023generating}.
More significantly, these methods are criticized by professional designers in terms of insufficient incorporation with domain knowledge, such as NCS~\cite{haard1981ncs,haard1996ncs1,haard1996ncs2} and color composition rules, and lack of support for user adjustments.
Through collaborative efforts and a formative study involving design practitioners, we gain a deep understanding of designers' needs for integrating AI within interior color design.
Subsequently, we introduce an LLM-based approach to achieve this objective.

\subsection{Human-AI Collaboration for Creative Design}
Design idea generation has traditionally been seen as an exclusively human endeavor, guided by numerous design guidelines and principles~\cite{hwang2022too}.
With the advances in AI, there has been a surge of interest in AI-supported creative design ideation~\cite{frich2018hci}, affecting broad-range applications spanning from art design \cite{cetinic2022understanding}, to sketch drawing~\cite{karimi2020creative}.
This concept harnesses the strengths of human intuition and AI's computational capabilities to navigate complex design spaces, ultimately enhancing human creativity and boosting productivity.
Generative models, including generative adversarial networks (GANs)~\cite{goodfellow2014generative}, variational auto-encoders (VAEs)~\cite{kingma2019introduction}, are now integral to AI-supported design processes (\eg,~\cite{nobari2021creativegan,cao2019ai}).
Simultaneously, diffusion models, renowned for their ability to produce rich and diverse samples, have found applications in areas such as artistic style transfer~\cite{zhang2023inversion}.
With its fast-evolving capabilities, AI has been playing an increasing role in human-AI collaboration for creative design~\cite{davis2013human, muller2022genaichi}.

However, AI-supported creative design has faced criticism for its perceived shortcomings in two crucial aspects: \emph{controllability} that pertains to the user's ability to influence, guide, or override the actions of AI, and \emph{rationality} that refers to the process aligning with established design principles and producing interpretable outputs.
In creative design, controllability is particularly vital since user intentions are often ambiguous and vaguely defined~\cite{cross1990nature, xu2014voyant}, necessitating flexible controls throughout the design process to ensure the outcomes align with the user's vision~\cite{wu2022ai}.
Many existing approaches are designed as end-to-end solutions, resulting in limited control over the generated outputs~\cite{duvsek2020evaluating}.
Potential solutions to address these issues encompass techniques like few-shot prompting~\cite{lin2021learning}, multi-step control~\cite{wu2022promptchainer, wu2022ai}, and interactive visual interface~\cite{strobelt2022interactive}.
Moreover, ensuring the rationality of results from these models poses a complex challenge.
Existing solutions primarily involve refining pre-trained models with domain-specific data~\cite{chen2023generating} and integrating human expertise through human-in-the-loop techniques~\cite{louie2020novice}.
Combining these approaches shows promise in generating designs that are both rational and reasonable.

In line with this prevailing research direction, our work harnesses LLMs to facilitate human-AI collaboration within the realm of creative interior color design.
Our primary emphasis centers on tackling the issue of controllability through an interactive visual interface and addressing rationality concerns by integrating color theory and design practices into the AI-assisted design process through prompts.

\subsection{Large Language Model with Domain Knowledge Integration}
LLMs have demonstrated remarkable potential in creative design.
However, a critical challenge lies in effectively incorporating design rationality and domain-specific knowledge~\cite{cui2023chatlaw, michalopoulos2020umlsbert, wang2023methods}.
Several methods have emerged to address the domain knowledge embedding challenge in LLMs.
One approach involves the integration of specialized external modules, assigning domain-specific tasks to dedicated models~\cite{peng2023check, shen2023hugginggpt}.
Alternatively, fine-tuning LLMs with domain-specific datasets offers a way to enhance their understanding of specific domains~\cite{dunn2022structured, xiao2023enhancing}.
Another approach is prompt engineering, which provides flexibility and control by using prompt templates and few-shot prompting techniques to guide LLM responses \cite{liu2023pre, wu2022promptchainer, wu2022ai}.
\re{This approach leverages `in-context learning'~\cite{brown2020language} to generate answers without altering the parameters of the pre-trained model.}
Lastly, retrieval-based methods introduce external knowledge bases through vector retrieval, enriching the LLM's knowledge base~\cite{shi2023replug}.

In our approach, we have chosen prompt engineering as the preferred method, primarily due to its flexibility, ease of control, and cost-effectiveness when compared to resource-intensive methods like fine-tuning \cite{liu2023pre}.
Inspired by PromptChainer~\cite{wu2022promptchainer} and AI Chains~\cite{wu2022ai}, we have deconstructed the color design process into steps aligned with common designer workflows.
This approach effectively addresses challenges related to interpretability and controllability often encountered when working with LLMs.
Instead of relying on traditional image generation models, we leverage LLMs endowed with reasoning capabilities to perform the design task.
Furthermore, we have incorporated user feedback at each stage of the design process, enabling iterative refinement until user satisfaction is achieved.
To overcome LLMs' performance limitations in specific domains, we encode multiple domain-specific knowledge elements related to interior color design into prompts integrated within our framework.
\section{\lowercase{\color{revision}}Formative Study}
\label{sec:preliminary}
We commenced a preliminary need-finding study to understand how designers are expecting AI assistance in the process of interior color design (Sect.~\ref{ssec:preliminary}).
\re{Before the study, we consulted with collaborating designers to understand} essential color theory (Sect.~\ref{ssec:domain_knowledge}) and the conventional interior color design workflow (Sect.~\ref{ssec:scenario}).

\subsection{Color Theory in Interior Design}
\label{ssec:domain_knowledge}
Color design professionals typically take into account color psychology, natural color system (NCS), and composition principles to craft emotionally resonant and visually harmonious spaces.

\subsubsection{Color Psychology}
Color schemes adopted in interior color design can reflect emotional responses.
Effective color design requires the understanding of user perceptions and the alignment of color schemes with space characteristics and user needs.
From a psychological standpoint, color impacts interior design in three primary dimensions:

\begin{itemize}
      \item \textbf{Temperature Sensation (warmth/coolness)}: Colors can be categorized as cold, warm, or neutral tones.
            Cold tones alleviate anxiety, warm tones heighten excitement, and neutral tones serve as transitions to alleviate visual fatigue\re{~\cite{haller2017colour, yildirim2011effects}}.
            For instance, hospitals often employ cooler tones to reduce anxiety~\re{\cite{dalke2006colour}}.
      \item \textbf{Spatial Perception (distance)}: Colors can influence the perceived size of interior spaces.
            Cool colors expand small spaces, while warm colors mitigate the sense of emptiness in oversized areas\re{~\cite{haller2017colour, al2017impact}}.
      \item \textbf{Weight Perception (lightness/heaviness)}: Colors can be classified as light or heavy, with purity and brightness positively correlating with lightness\re{~\cite{alexander1976influence}}.
            This dimension affects interior color composition, such as using heavier colors for the floor to create a sense of stability.
\end{itemize}

These three dimensions of color psychology are interconnected, requiring designers to consider their mutual influence in practice.
For example, in a narrow bedroom where users seek a cozy experience, designers must balance conflicting demands for warmth and spatial extension.

\subsubsection{Natural Color System (NCS)}
NCS is commonly used in interior design due to its alignment with fundamental color perception and its compatibility with cognitive and linguistic aspects of color\re{~\cite{haard1981ncs,haard1996ncs1,haard1996ncs2}}, as follows:

\begin{itemize}
      \item \textbf{Hue}: There are six base hues, including white (W), black (S), yellow (Y), red (R), blue (B), and green (G).
      \item \textbf{Blackness}: Blackness represents how bright or dark a color is within the NCS system.
      \item \textbf{Chromaticness}: Chromaticness refers to the saturation or intensity of a color.
\end{itemize}

\subsubsection{Color Composition}
In interior color design, designers typically identify the primary color tone based on NCS and consider effective color composition, to enhance visual coherence and prevent color imbalances\re{~\cite{haller2017colour}}.
For instance, consider a warm-toned bedroom, the right composition of orange colors can create a cozy atmosphere, whilst excessive orange colors can cause boredom, anxiety, and visual fatigue\re{~\cite{kwallek1988effects}}.
Typical color compositions are set as follows:
\begin{itemize}
      \item \textbf{Primary Color (60\% to 70\%)}: The primary color determines the interior style and atmosphere, which are often applied to walls, floors, and furnishings.
      \item \textbf{Secondary Color (20\% to 30\%)}: The secondary color is used to enhance color harmony, typically through the use of similar or contrasting or complementary color combinations with the primary color.
      \item \textbf{Accent Color (5\% to 10\%)}: Accent colors are typically high-contrast colors, which are used to promote a balanced, rich interior color scheme through careful balancing.
\end{itemize}

\begin{figure}[htb]
      \centering
      \includegraphics[width=0.98\linewidth]{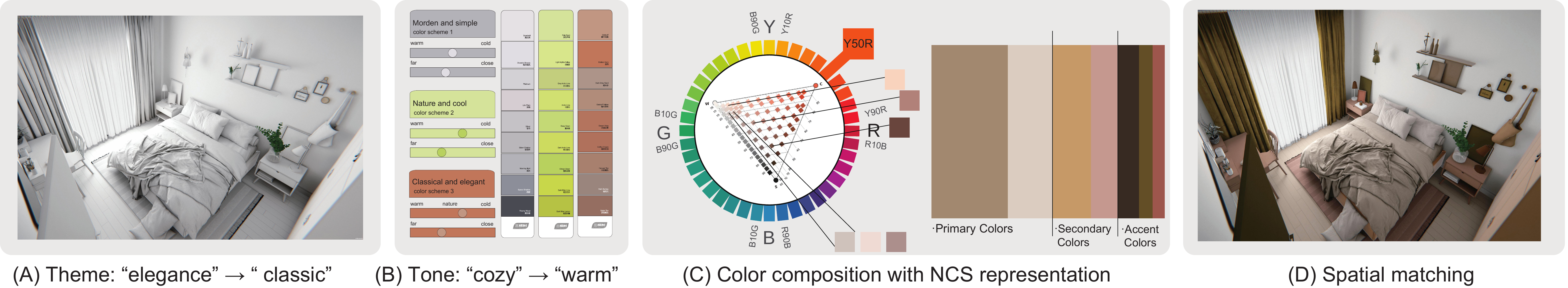}
      \caption{Interior color design workflow. 1) Concept design step associates design concepts with styles (A) \& tones (B). 2) Color composition step decides primary, secondary, and accent colors (C). 3) Spatial matching step assigns colors to interior elements (D).}
      \vspace{-4mm}
      \label{fig:workflow}
\end{figure}

\subsection{Design Practice}
\label{ssec:scenario}
To elucidate interior color design practices, we commence with an illustrative scenario centered on bedroom design, as shown in Figure~\ref{fig:workflow}.
The workflow can be divided into three steps:

\begin{enumerate}
      \item \textbf{Concept Design.}
            In this step, designers summarize the interior color theme based on the user's needs and in line with the function and context of the interior space.
            For instance, if a cozy and elegant bedroom is desired, a designer will associate \q{elegance} to \q{classical} which is a common terminology in interior color design, as shown in Figure~\ref{fig:workflow} (A).
            Consequently, warm tones are often chosen to create a cozy ambiance, as shown in Figure~\ref{fig:workflow} (B).

      \item \textbf{Color Composition.}
            The second step involves semantic-to-color conversion that constructs a color scheme based on the chosen themes and tones.
            To control the presence of large areas with highly chromatic colors, designers need to manage the proportions of the primary, secondary, and accent colors.
            The primary color's hue change should be limited to 15 degrees for brightness adjustment.
            Secondary colors have three choices: similar colors, contrasting colors, or complementary colors.
            For example, classical styles often utilize high-purity warm colors like cream, beige, and tan to create a serious but not overpowering tone, as depicted in Figure~\ref{fig:workflow} (C).
            In this way, a more harmonious color scheme that aligns with the \q{classical} theme is derived.

      \item \textbf{Spatial Matching.}
            In the third step, designers apply the color scheme to the interior elements and evaluate the visual effects of color matching.
            Designers make precise adjustments to the color composition ratios, transitioning from overall control to finer details.
            For instance, once the selected bedroom color scheme successfully conveys the desired \q{warm and elegant} atmosphere, designers explore opportunities to enhance interior color richness by incorporating creative colors that reflect the user's personality, as shown in Figure~\ref{fig:workflow} (D).
            This step plays a pivotal role in preserving the integrity and harmony of the interior color scheme, preventing visual fatigue, and elevating the overall artistry and appeal of the design.
\end{enumerate}

\subsection{Preliminary Study: Human-AI Collaboration in Interior Color Design}
\label{ssec:preliminary}

\subsubsection{Study Design}
We recruited six participants for the preliminary study, including three junior designers (P1, P2, P3) with less than 5 years of experience and three senior designers (P4, P5, P6) with more than 5 years of experience in color design.
All the participants are familiar with the color design principles.
Semi-structured interviews served as the primary data collection method, aimed at eliciting insights into the designers' color workflow, their current interactions with color design tools, and their expectations for AI assistance.
The interview was conducted in a quiet room and lasted for about 1 hour for each participant.
Each interview was audio-recorded for subsequent analysis.

\subsubsection{Findings}
To analyze the qualitative data gleaned from the interviews, we transcribed the recorded conversations and structured the data according to the designers' needs for AI assistance, as follows:

\textbf{Design Workflow:}
Traditional interior color design companies typically maintain a database containing color schemes associated with various design styles.
These resources facilitate the generation of low-fidelity prototypes, improving communication with clients to ensure alignment with their objectives.
Junior designers delineate a multi-step design workflow involving client consultations, color research, sample collection, iterative design processes, client feedback, and finalization.
They encounter difficulties in translating client intentions into appropriate color concepts.
In contrast, experienced designers follow a similar workflow but underscore their adeptness in efficiently grasping client insights.
As mentioned by P5, \q{eighty percent of an interior color designer's work time is spent communicating with the client over and over again, rather than tweaking the design itself}.
Even experienced designers dedicate a significant portion of their time to client communication and understanding their preferences, underscoring the necessity for streamlined tools to bolster the interpretation of client intentions effectively.
This underscores the paramount importance of understanding and translating \re{vague intent from client} within the entire design process.

\textbf{Collaboration with Computer-Aided Design Tools \& Expectations for AI Assistance:}
In interior color design, designers commonly employ computer-aided design tools primarily for rendering results, with the actual color design being executed by the designers themselves.
In recognition of recent advances in AIGC technologies, junior designers express a desire to access reasonable design samples to serve as guiding references for their creative endeavors.
In contrast, experienced designers employ predominantly manual methods without the involvement of image generation tools.
They leverage their expertise to manually craft color schemes, observing that AI-generated outcomes often lack reasonability and originality.
Moreover, the process of modifying these generated results proves to be arduous and time-consuming, as it necessitates iterative adjustments to the prompts through a trial-and-error process.
As participant P4 aptly remarked, \q{when I have a clear vision already, I prefer executing it myself rather than instructing the AI, as guiding the AI can be a time-intensive endeavor.}
The central challenge they encounter pertains to achieving controllability and making precise adjustments effectively to the AI-generated outputs.

From the preliminary study, we delineate three primary design requirements:

\begin{itemize}
      \item \textbf{DR1: Enhancing Comprehension of Users' Intent.}
            The system is expected to adeptly grasp and translate the intent of designers, bridging it to the design lexicon.
            The goal is to streamline the process for designers, enabling them to meet \re{their clients'} needs with minimal exertion.

      \item \textbf{DR2: Achieving Practical Results.}
            The emphasis is on ensuring that the tool effectively correlates the user's intent with the final design.
            The output should resonate with the designers' aspirations, comply with design rationales, and encapsulate an element of creativity and innovation.

      \item \textbf{DR3: Ensuring a Controllable Design Process.}
            The design process should grant ample autonomy to designers.
            They should have the liberty to control each step, equipped with tools for meticulous adjustments.
            This seeks to amalgamate the precision of automation with the uniqueness of a personal touch.
\end{itemize}

\section{System Overview}
\label{sec:overview}

In this study, our aim is to create an LLM-assisted tool \re{to assist designers in} interior color design.
The tool is designed to seamlessly integrate with user intentions while adhering to established design principles.
The LLM is anticipated to: 1) articulate the curated design principles through well-defined prompts;
2) consume user-desired ambiances or styles that are often expressed in fuzzy and ambiguous natural language;
and 3) assign colors to interior elements based on their spatial arrangements and sizes.

\begin{figure}[t]
      \centering
      \includegraphics[width=0.99\linewidth]{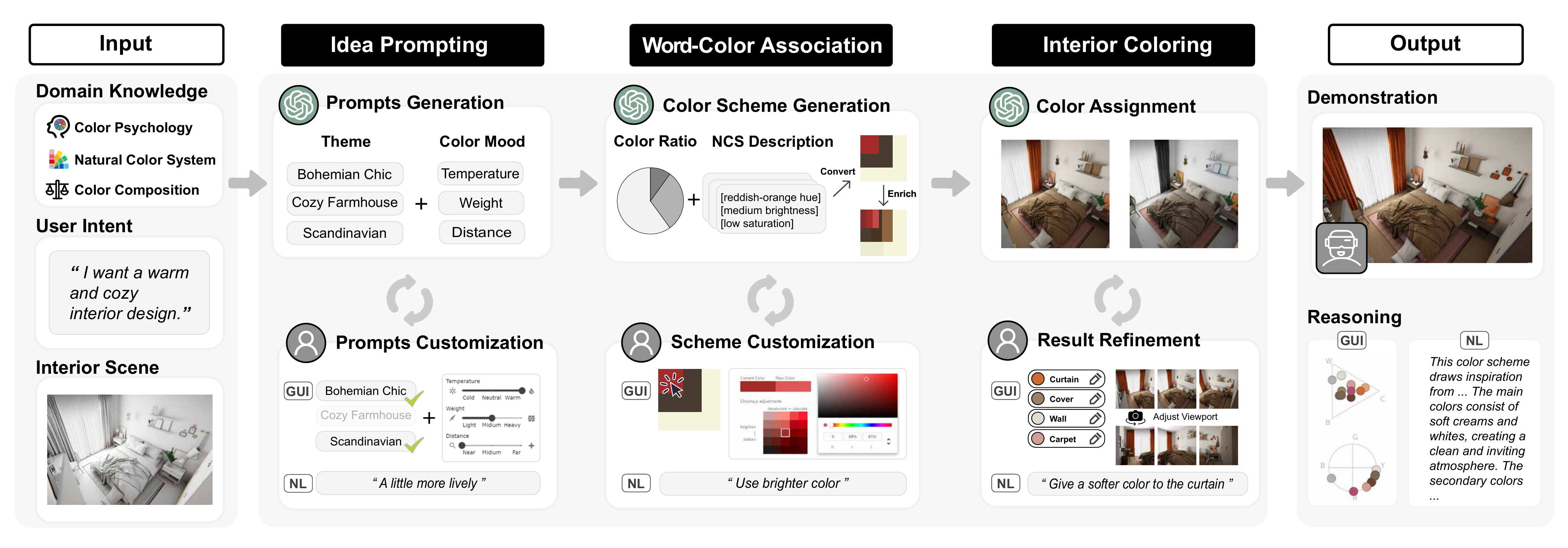}
      \vspace{-5mm}
      \caption{\tool includes three stages:
      1) Idea Prompting: Prompts Generation and 
       Prompt Customization; 2) Word-Color Association: Color Scheme Generation and Color Scheme Customization; 3) Interior Coloring: Color Assignment and Result Refinement. 
       }
      \vspace{-4mm}
      \label{fig:overview}
\end{figure}

To attain the goals, we design \tool, an innovative framework with three integral stages: \textit{Idea Prompting}, \textit{Word-Color Association} and \textit{Interior Coloring}, as illustrated in Figure~\ref{fig:overview}.
These stages closely follow the conventional workflow of interior color design, thus enhancing the interpretability and controllability of the generation process, and ensuring adherence to design principles.
Each stage combines \llmop{LLM Operations} with corresponding \interaction{User Interactions}.

\begin{enumerate}
      \item \emph{Idea Prompting} (Sect. \ref{ssec:prompting}).
            In the first stage, we aim to translate user intentions into a set of well-defined prompts, which are then used to guide the subsequent process.
            In particular, \llmop{Prompts Generation}
            leverages LLMs to decompose user's input into a set of domain-oriented prompts, corresponding to design themes and color mood attributes.
            In \interaction{Prompt Customization}, we allow designers to modify and customize the generated prompts, to meet the diversity of personalized preferences for the same concept.

      \item \emph{Word-Color Association} (Sect. \ref{ssec:text_color}).
            In the second stage, we aim to transfer conceptual color prompts to definitive color schemes, keeping semantic and stylistic coherence.
            \llmop{Color Scheme Generation} leverages LLMs to generate NCS color descriptions of 3-color schemes considering the color contraction principle, and then add color diversity through variations in saturation and lightness.
            \interaction{Color Scheme Customization} enables designers to modify the generated color schemes as if the suggested color does not fit their design concepts.

      \item \emph{Interior Coloring} (Sect. \ref{ssec:interior_coloring}).
            In the third stage, we color the interior scene using the generated color scheme, test how well the scheme fits the interior space and refine the result if necessary.
            \llmop{Color Assignment} leverages LLMs to assign one of the chosen color schemes from the previous step to each interior element.
            \interaction{Result Refinement} allows designers to adjust the color for each interior element.
            Designers can also switch their viewport between first-person and third-person perspective and rotate the camera to better validate the result.
\end{enumerate}

The coloring results are demonstrated in the 3D scene through Virtual Reality (VR), in which designers can immersively experience the design.
Besides, result reasoning is supported by visual analysis of used colors, and natural language descriptions for the design ideas.
\section{C2Ideas}
\label{sec:methods}
This section presents a detailed implementation of \emph{C2Ideas}, a structured three-stage framework following the idea of chain-of-thought (CoT) to harness the reasoning capabilities of LLMs for interior color design (\textbf{\re{DR1}}).
\re{The implementation aligns with the interior color design workflow, with the outcome of one stage serving as the input for the next, ensuring a coherent and interconnected process. }
Each stage integrates an \llmop{LLM Operation} with an associated \interaction{User Interaction}.
For \llmop{LLM Operation}, we design well-crafted prompt templates following the principle of clarity and specificity, ensuring that the LLMs align seamlessly with the specific design rationales (\textbf{\re{DR2}}).
For \interaction{User Interaction}, we incorporate natural language with a graphical interface to facilitate nuanced customization with graphical interaction and general user needs with text inputs (\textbf{\re{DR3}}).
In the following, we introduce details of each stage (Sects.~\ref{ssec:prompting} $-$ \ref{ssec:interior_coloring}), as well as the interactive interface (Sect.~\ref{ssec:system_design}) \re{and implementation details (Sect.~\ref{ssec:details})} of \emph{C2Ideas}.

\subsection{Idea Prompting}
\label{ssec:prompting}
\subsubsection{Prompt Generation}
\label{sssec:prompt_generation}
\begin{figure}[htb]
    \centering
    \includegraphics[width=0.88\linewidth]{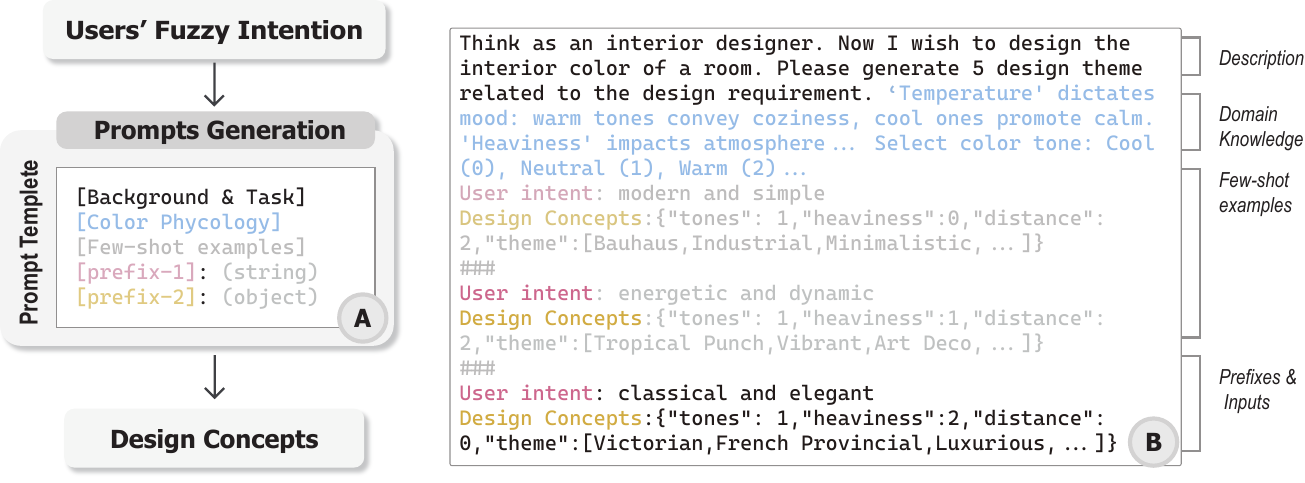}
    \caption{Idea prompting stage.
        (A) the prompt template, (B) an example of the prompt that maps the user's input to the design concepts, considering the domain knowledge of color psychology theory.
    }
    \vspace{-4mm}
    \label{fig:prompt_generation}
\end{figure}

The first step is to translate users' high-level design intents into a set of domain-oriented prompts \re{that are composed of professional design themes and color mood attributes, as suggested by our collaborating designers and in line with previous works~\cite{lin2020c3, solah2022mood}.}
We design a well-crafted prompt template to guide the generation of prompts, composed of four components as shown in Figure~\ref{fig:prompt_generation} (A).

\begin{itemize}
    \item \textit{Background \& Task Description}:
          The background and task description are used to provide the context of the design task.
          This includes perspective-taking prompts \cite{park2023generative} such as \q{think as an interior designer}, and task specification prompts such as \q{please generate 5 design themes related to the given design requirement}.
    \item \textit{Domain Knowledge}:
          Completing a creative design task requires following design rationales, which are often domain-specific.
          We transfer the color psychology theory from the preliminary study into a series of well-defined prompts, which is interpretable for LLMs.
          This mainly includes domain knowledge on color psychology, such as \q{warm tones convey coziness} and \q{cool ones promote calm},
          \re{and also attribute specifications such as color mood that are divided into three categories of \q{cool (0), neutral (1), and warm (2)}.}
    \item \textit{Few-shot Examples}:
          As proved by many works, LLMs are capable of generating high-quality results given few-shot examples~\cite{liu2023pre}.
          \re{In interior color design, there are professional design themes and their corresponding color moods that are hard to summarize as high-level design rules, such as the mapping from user intent of \q{modern and simple} to design concepts of \q{Bauhaus} and \q{Minimalistic}.
          As such, these mappings cannot be included in the Domain Knowledge component.
          To address the issue, we complement the domain knowledge with few-shot examples by providing pairs of user's vague input and the corresponding design concepts, as shown in Figure~\ref{fig:prompt_generation} (B).}
          Specifically, we include design themes and three important attributes that can reflect the physical impact of color, namely \textit{tones}, \textit{distance}, and \textit{heaviness}.
          Each criterion is quantified to 3 degrees, such as \q{warm, neutral, cool} for tones, \q{close, medium, far} for distance, and \q{light, medium, dark} for heaviness.
          To ensure accuracy and consistent representation, we quantify and convert the natural language description into a specific degree.
          \re{In addition to facilitating high-quality result generation, few-shot examples can guide LLMs to generate structured data, which is effective for system implementation.}
    \item \textit{Prefix \& Inputs}:
          These can act as specific cues or triggers to further guide the LLM-generated design concepts based on the user's input intent.
          Specifically, user intents are expressed in strings, while the design concepts are represented in a JSON format to convey multi-perspective domain knowledge.
\end{itemize}

An example of the generated prompt is shown in Figure~\ref{fig:prompt_generation} (B).
Here the user wants an \q{energetic and dynamic} interior color design, and \llmop{Prompts Generation} transfer it into design concepts, including themes such as \q{Tropical Punch, Vibrant, Art Deco}, and color mood attributes including \q{neutral tones}, \q{far}, and \q{medium heaviness}.

\subsubsection{Prompt Customization}
\label{ssec:prompt_customization}

\tool supports both graphical interface and natural language approaches for prompt customization.
Users can select appropriate design concepts from the generated ones using the graphical interface as shown in Figure~\ref{fig:interface} (B1).
The design themes are presented as tags, and the color mood attributes, including tones, distance, and heaviness are presented as sliders.
Users can select the tags they want to include, as well as adjust the degree of each color mood attribute.
This functionality enables personalized choice of color themes and color moods, based on users' prior knowledge and preference.
For example, for \q{cozy}, some people will think that it needs a large area of warm tones, while others may think that neutral colors such as cream are more in line with their perceptions.

The graphical interface approach is more suitable for users with clear intentions.
We also provide a natural language approach for users with vague intentions.
It is common that the user's input has multiple meanings, and the user's true intention is only part of it.
For instance, the concept \q{modern} can be interpreted as \q{modern minimalism} or \q{dynamic and bold}, which are two different design styles.
The user can adjust the generated prompts by supplementing the input with additional information, such as adding \q{modern simple} to indicate the \q{modern minimalism} style.

\subsection{Word-Color Association}
\label{ssec:text_color}

\subsubsection{Color Scheme Generation}
\label{ssec:palette_generation}

\begin{figure}[t]
    \centering
    \includegraphics[width=0.975\linewidth]{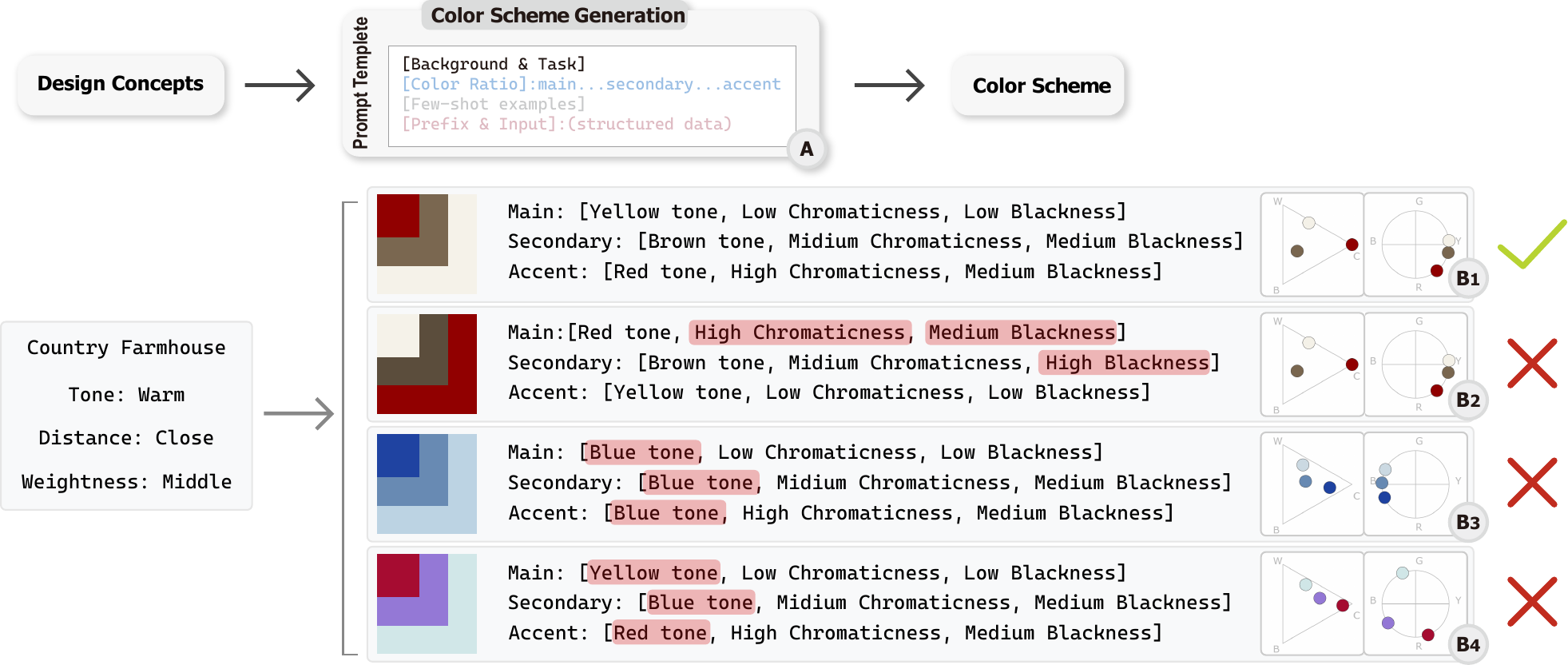}
    \vspace{-2mm}
    \caption{Word-color association stage. Given a prompt template (A), the LLM for Word-Color Association returns a positive result (B1) with correct reasoning whilst rejecting three negative examples violating color construction theory (B2 - B4).}
    \vspace{-2mm}
    \label{fig:Step2}
\end{figure}

This stage processes the refined design concepts derived from the \llmop{Prompt Generation} stage into concrete color schemes.
We design a domain-oriented LLM with prompt templates encompassing design rationale of background \& task, color composition ratios, few-shot examples, and prefix with input, as illustrated in Figure~\ref{fig:Step2} (A).
Examples of these concepts are descriptors such as \q{Tropical Punch} and \q{Vibrant}, in addition to color mood attributes like \q{neutral tones}, \q{far}, and \q{medium heaviness}.
\re{LLM translates these concepts into a coherent color scheme comprising three colors.
    This translation process leverages the LLM's ability to map semantically similar words, which is influenced by both the model's pre-training and the specific context of the prompt.
    In the model's embedding space, words with similar meanings are situated in close proximity, enabling the effective identification of relevant color associations.}
To streamline the design process, we include a few-shot examples, providing a benchmark of ideal color schemes.
The scheme is then enriched by adjusting variations in saturation and lightness.
\re{
    To enhance LLM's interpretation of color and ensure the coherence of color schemes with concept descriptors, we incorporate the foundational design principles of NCS representation and color composition rules rather than HEX code or RGB values that lack semantic meaning.
}
NCS aligns with human visual perceptual processes, which also employ common color descriptions for colors that are easily understood by the model.
Every color in the color scheme adopts the NCS representation, accounting for hue, chromaticness, and blackness.
In addition, we employ structured composition to delineate each color scheme, classifying colors into \q{dominant}, \q{secondary}, and \q{accent} categories.

With the assistance of LLMs, design concepts are matched with color design domain knowledge to ensure semantic and stylistic cohesion.
\re{As exemplified in Figure~\ref{fig:Step2} (B1), the provided design concepts associate the \q{warm tone} to \q{red hue}, rather than \q{blue} or \q{green}, which are far from the term \q{warm} in the embedding space.}
Similarly, the brown and woody hues correspond to \q{country farmhouse}, complemented by a light and desaturated primary color of a similar hue.
The composition of colors adheres to specific design guidelines \re{specified in "color ratio" shown in the prompt template.}
In contrast, the scheme in Figure~\ref{fig:Step2} (B2) is deemed inappropriate because its dominant color is excessively dark and saturated, making it unsuitable for expansive spaces.
Similarly, the scheme in Figure~\ref{fig:Step2} (B3) is rejected due to its hue deviating from the stipulated tone, potentially inducing contrasting emotions.
Finally, as depicted in Figure~\ref{fig:Step2} (B4), the generated result appears overly diverse, making it challenging to establish a distinct emotional connection.
Consequently, this diversity renders it unsuitable for interior color design purposes.

\subsubsection{Color Scheme Customization}
\label{sssec:palette_customization}

Upon generating a color scheme that maps design concepts to specific hues, we offer an interactive customization interface, as in Figure~\ref{fig:interface} (B2).
This GUI provides users with the capacity to modify scheme colors utilizing a responsive color picker.
As users navigate through these adjustments, the scheme updates contemporaneously, granting immediate feedback.
For instance, if the user finds that the primary and secondary colors appear overly analogous, users have the liberty to refine attributes like saturation and lightness to establish clear distinctions.
An integral feature is the \q{lock} mechanism, which permits users to freeze certain color selections, ensuring only the desired hues undergo alterations.
Understanding the potential complexities of manual color adjustments, we've complemented the GUI with a natural language interaction feature.
For instance, if a user finds the generated scheme overly dark, a straightforward instruction such as \q{make the color scheme brighter} prompts the system to adjust the luminosity consistently across the scheme.
This natural language capability offers a quick and intuitive alternative for users, emphasizing simplicity and user-friendliness.

\subsection{Interior Coloring}
\label{ssec:interior_coloring}

\subsubsection{Color Assignment}
\label{sssec:color_assignment}

\begin{figure}[t]
    \centering
    \includegraphics[width=0.9\linewidth]{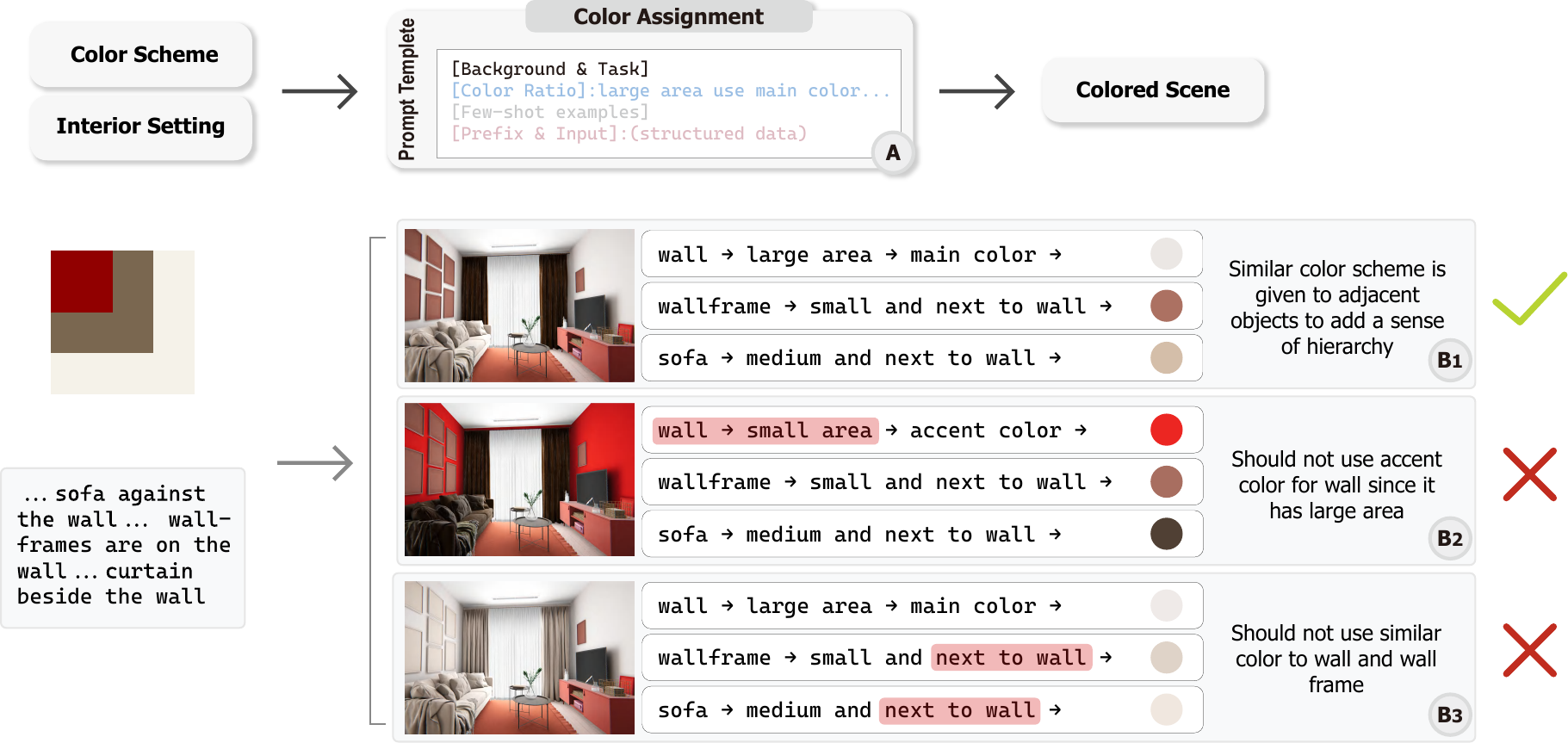}
    \caption{Interior coloring stage. The color scheme and interior settings are formatted into the prompt template (A). The proper design result (B1) considers sizes and layouts of interior elements, whilst rejecting negative examples that violate color compositions (B2, B3).}
    \vspace{-4mm}
    \label{fig:Step3}
\end{figure}

This stage takes a color scheme and \re{a textual description of the interior scene} as inputs and assigns colors to the corresponding interior elements.
We take into account the sizes of interior elements in relation to the composed colors of the color scheme, and also consider the impact of the spatial arrangement of these elements on the perception of the color assignment.
\re{For scene interpretation, we utilize manually crafted textual descriptions that encompass all furniture, fixtures, and others, along with their size attributes and layout arrangements, such as \q{In the center of the room against the right wall is a bed with a cover and pillows near the cover}.
    These descriptions provide detailed spatial context, enabling the LLM to identify and associate furniture items like \q{bed}, \q{cover}, and \q{pillows} with suitable color schemes, based on their sizes and arrangements.
    The LLM interprets \q{bed} as a large item and \q{pillows} as smaller, using this size inference in combination with prompt templates to establish connections between furniture size and composition of main, secondary, and accent colors.}
The assignment results are structured with color codes and rendered into a 3D scene, which is more intuitive for users to evaluate how the color scheme is applied to the interior.
Reasoning is also provided to facilitate the interpretability of the results.

Figure~\ref{fig:Step3} (A) shows an example of the structured input for the LLM to assign colors to each interior element and fixture.
Three alternative assignments are produced, as in Figure~\ref{fig:Step3} (B1 - B3).
\re{Similar with the previous stage, we take design rules of color ratio as instruction for the LLM to generate proper color assignments, and those that violate these rules are avoided, as shown in Figure~\ref{fig:Step3} (B2, B3).
}
The design in Figure~\ref{fig:Step3} (B1) is acceptable since the results are in alignment with the element sizes and the color composition theory.
The walls, for example, occupy a large area and, therefore, correspond to the main color of the color scheme, using a cream color with high brightness and low saturation.
The decorative wall hangings use a similar hue of red, which contrasts with the wall color while keeping the overall ambiance in line with \q{warm and cozy}.
In contrast, the wall in Figure~\ref{fig:Step3} (B2) is mapped to accent color with a saturated red, which will keep the residents stimulated and easy to get tired.
In Figure~\ref{fig:Step3} (B3), the decoration frames on the wall are wrongly mapped to the secondary color, which is similar to the color of the wall, making it hard to distinguish the wall from the frame.
This will make the whole wall become monotonous in color, whilst contrast colors are often used to increase the depth of the space for elements such as the wall and the decoration frame.

\begin{figure}[t]
    \centering
    \includegraphics[width=0.99\linewidth]{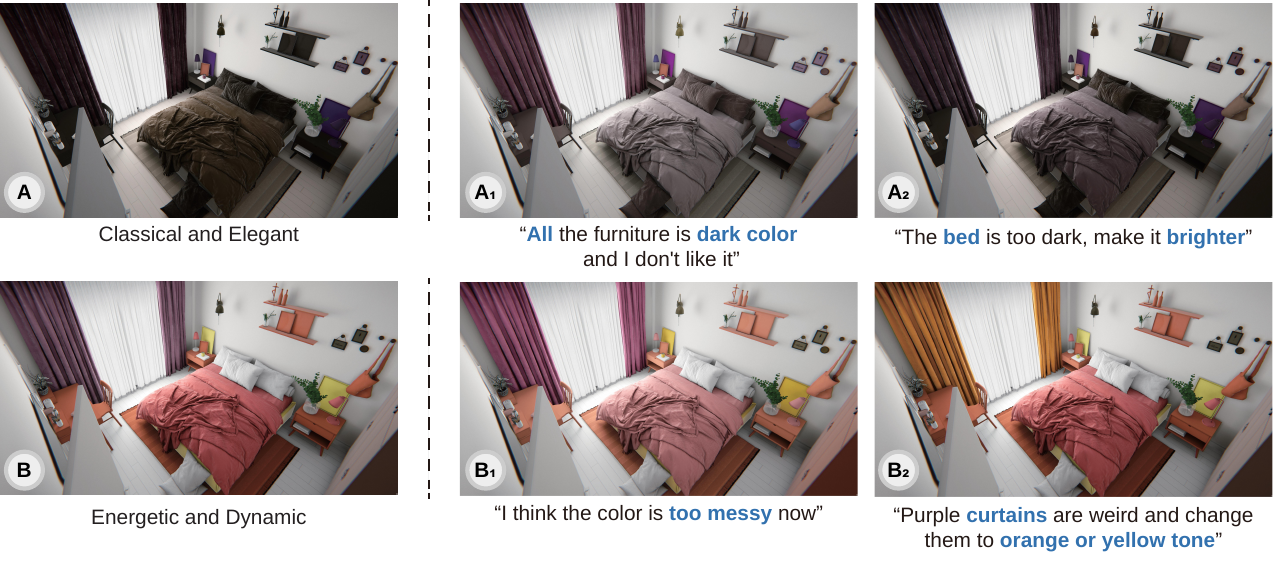}
    \vspace{-3mm}
    \caption{
        Examples of using natural language interaction to enhance color design.
        (A) "Classical and Elegant" bedroom's initial design: (A1) refinement based on adjusting overall darkness, and (A2) effect of altering the bed's color.
        (B) Adjustments for an "Energetic and Dynamic" bedroom: (B1) tidying up the color scheme, and (B2) specifying the curtain's color as orange.
    }
    \vspace{-3mm}
    \label{fig:Case-Improve}
\end{figure}

\subsubsection{Result Refinement}
\label{sssec:result_refinement}

\begin{figure}[t]
    \centering
    \includegraphics[width=0.99\linewidth]{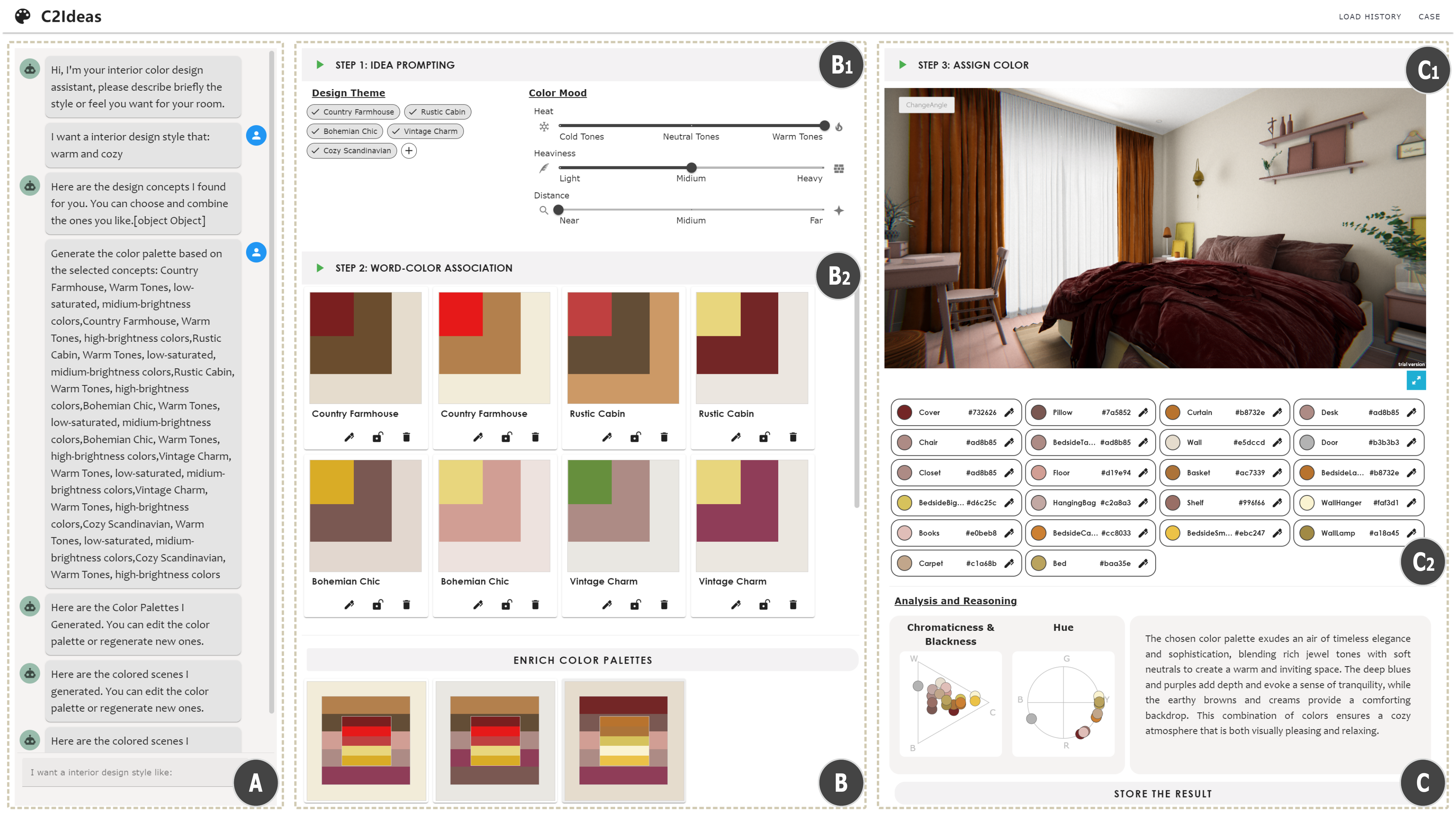}
    \vspace{-2mm}
    \caption{Interface of our system including (A) Conversation View for users to input their initial design intents and customizations expressed in natural language, (B) Color Design View for users to customize the intermediate results generated by the LLMs, and (C) Result View for users to evaluate the results with reasoning and refine the color assignment.}
    \vspace{-3mm}
    \label{fig:interface}
\end{figure}

The automated process of \llmop{Color Assignment} is complemented with a GUI for user evaluation and result refinement, as depicted in Figure~\ref{fig:interface} (C).
Recognizing that the viewport can influence the perceived ambiance, we've incorporated functionalities for viewport adjustments.
Users can toggle between first and third perspectives and modify the camera angle using the mouse.
Users have the capability to modify the colors of specific elements.
For instance, while appreciating the broader color scheme of an interior space, a user might find the orange hue of the curtain disagreeable and prefer a purple tone.
The system allows such precise changes, updating the element's color in the 3D scene in real-time.
On the other hand, granular changes presuppose users have specific intentions.
At times, users might harbor general feelings of dissatisfaction without pinpoint clarity on adjustments.
They might, for example, feel a design \q{use messy color} without specifics on how to improve it.
To accommodate such general sentiments, we have integrated natural language interactions, enabling users to convey varying degrees of specificity in their feedback, from overarching ambiance shifts to elemental color tweaks, as demonstrated in Figure~\ref{fig:Case-Improve}.
\re{The user's instruction and existing color scheme are formatted as prompts for the LLM to generate the refined color scheme.
Enhanced by NCS representation, the LLM interprets the user's instruction and existing color scheme, such as mapping \q{brighter} to \q{blackness} attribute and \q{orange} to \q{hue} attribute.
}
For instance, the left scene shows initial results automatically generated by the \llmop{Color Assignment} model.
The user may wish to alter the design style, such as requesting that \q{all the furniture is dark color and I don't like it}.
The entire scene will be updated with new furniture colors while maintaining the overall color scheme's \q{classical and elegant} appearance.

\subsection{Interactive User Interface}
\label{ssec:system_design}

We develop an interactive system to complement the automated interior color designs by the LLMs, as illustrated in Figure~\ref{fig:interface}.
There are three views in the web interface:

\begin{enumerate}
    \item \textbf{Conversation View} (Figure~\ref{fig:interface} (A)).
          The view allows users to enter their own design intention, such as some adjectives or keywords of the desired color style in the dialog box.
          The LLMs will automatically generate corresponding design themes and color moods, generate appropriate word-color associations, and assign colors to the interior elements.
          The view also allows users to view the history of conversations and input further refinement instructions.

    \item \textbf{Color Design View} (Figure~\ref{fig:interface} (B)).
          The view allows users to see the intermediate results of color design and can adjust the results in each stage respectively.
          The first step is called Idea Prompting, which contains two parts: Design Theme and Color Mood. Users can select the design theme tags they want to include, as well as adjust color properties such as Temperature, Heaviness, and Distance, and users can adjust the warm and cool tones of the color (Figure~\ref{fig:interface} (B1)). In the second step of Word-Color Association, a series of initially generated Word-Color Association schemes are presented to the user. Users can select the most appropriate color scheme based on the degree of association between the colors and the corresponding words  (Figure~\ref{fig:interface} (B2)).

    \item \textbf{Result View} (Figure~\ref{fig:interface} (C)).
          The view allows users to see the spatial color scheme displayed in the upper right part as a concrete internal coloring result. Meanwhile, all the colorable objects in the scene and their corresponding hexadecimal color information are listed below for users' reference (Figure~\ref{fig:interface} (C1)). For a more professional analysis of the color composition, we provide a Chromaticness and Blackness chart, as well as a Hue Distribution chart. Finally, we also explain the reasons for using the above color scheme to help users better understand the color information generated in the previous section, and users can also store the current results for later use (Figure~\ref{fig:interface} (C2)).
\end{enumerate}

\subsection{\re{Implementation Details}}\label{ssec:details}
    \re{This section details our system's implementation.
    The system backend is built using Flask integrated with OpenAI's GPT-3.5 Turbo API.
    The architecture is modular, with distinct functions for each LLM operation.
    These functions execute sequentially, with each stage's output becoming the next's input.
    For each stage, we use few-shot examples that cover common design styles provided by our collaborating designers.
    All the examples and outputs are in a structured format, which makes it easy for the LLMs to interpret.
    Following conventions in few-shot prompting, we provide 3-5 examples in each stage.
    Specifically, in the \llmop{Prompt Generation} stage, five examples are provided spanning various design themes and color moods; in the \llmop{Color Scheme Generation} stage, another five examples of color schemes manually designed by interior designer are provided using existing color palette in the company; in the \llmop{Color Assignment} stage, three examples are provided for each interior scene.
    These examples strike a balance between providing sufficient context and maintaining brevity for efficient model performance.
    Token numbers are limited to 4096 for each stage, which is the maximum number of tokens that can be processed by the GPT-3.5 Turbo API.
    Temperature is set to the default value of 0.7 for all stages.
    The example prompts are presented in the Supplementary Material. 
}

\re{The interface is a web-based application crafted using Vue.js, with graphical components rendered through WebGL.
    To enhance the immersive experience of the design results, we have created a VR component using the Unity platform.
    This component optimizes the scene for both WebGL and VR displays.
    Additionally, we have tackled challenges such as perspective adjustments to ensure an optimal viewing experience for users.
}
\section{Evaluation}
\label{sec:evaluation}

\subsection{User Study}

\begin{figure}[t]
    \centering
    \includegraphics[width=0.95\linewidth]{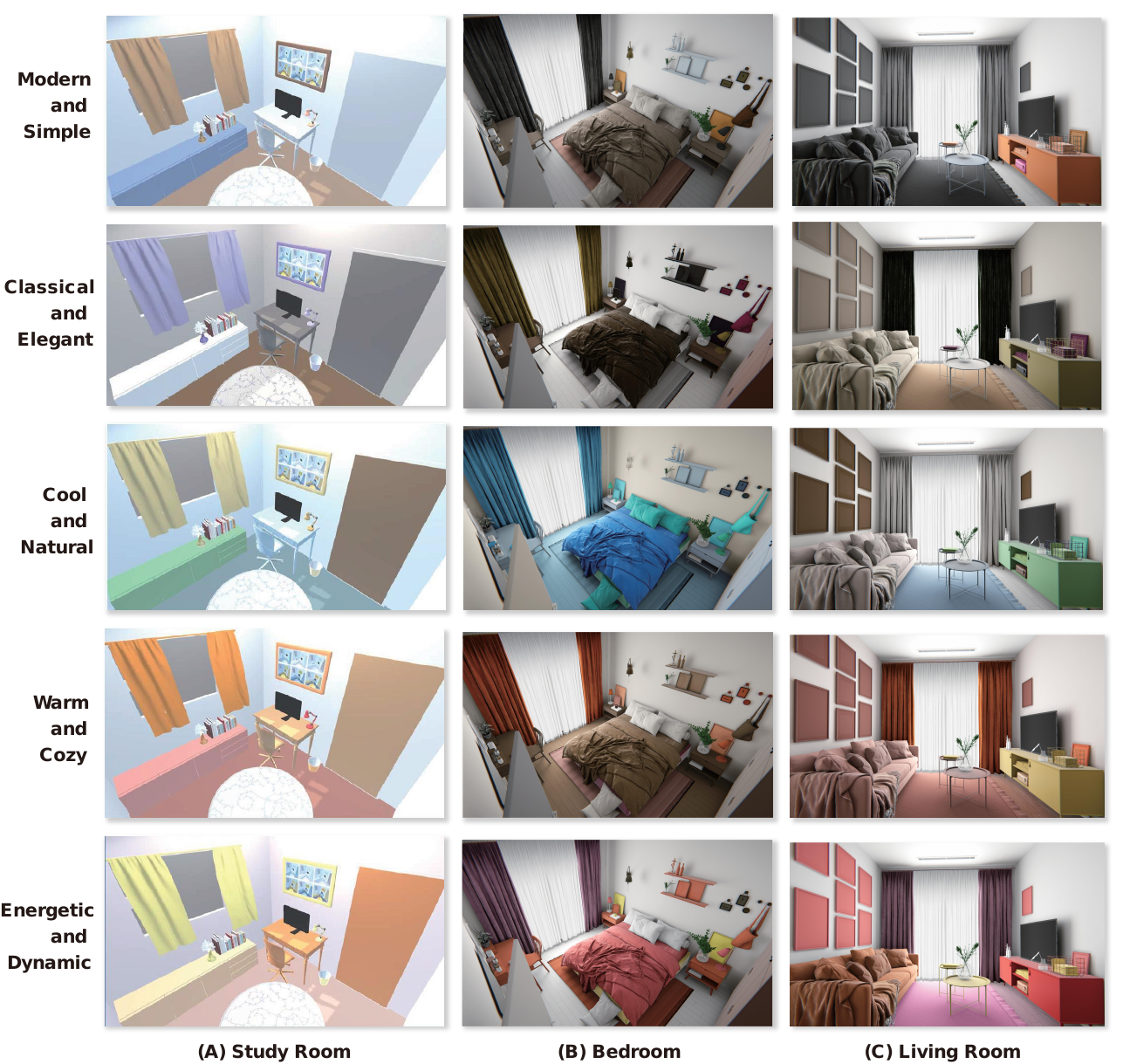}
    \caption{Examples of colored scenes with different room settings and styles, including simple scene (A) study room, and complex scene (B) bedroom and (C) living room, \re{by \tool}.}
    \vspace{-4mm}
    \label{fig:cases}
\end{figure}

\begin{figure}[t]
    \centering
    \includegraphics[width=0.99\linewidth]{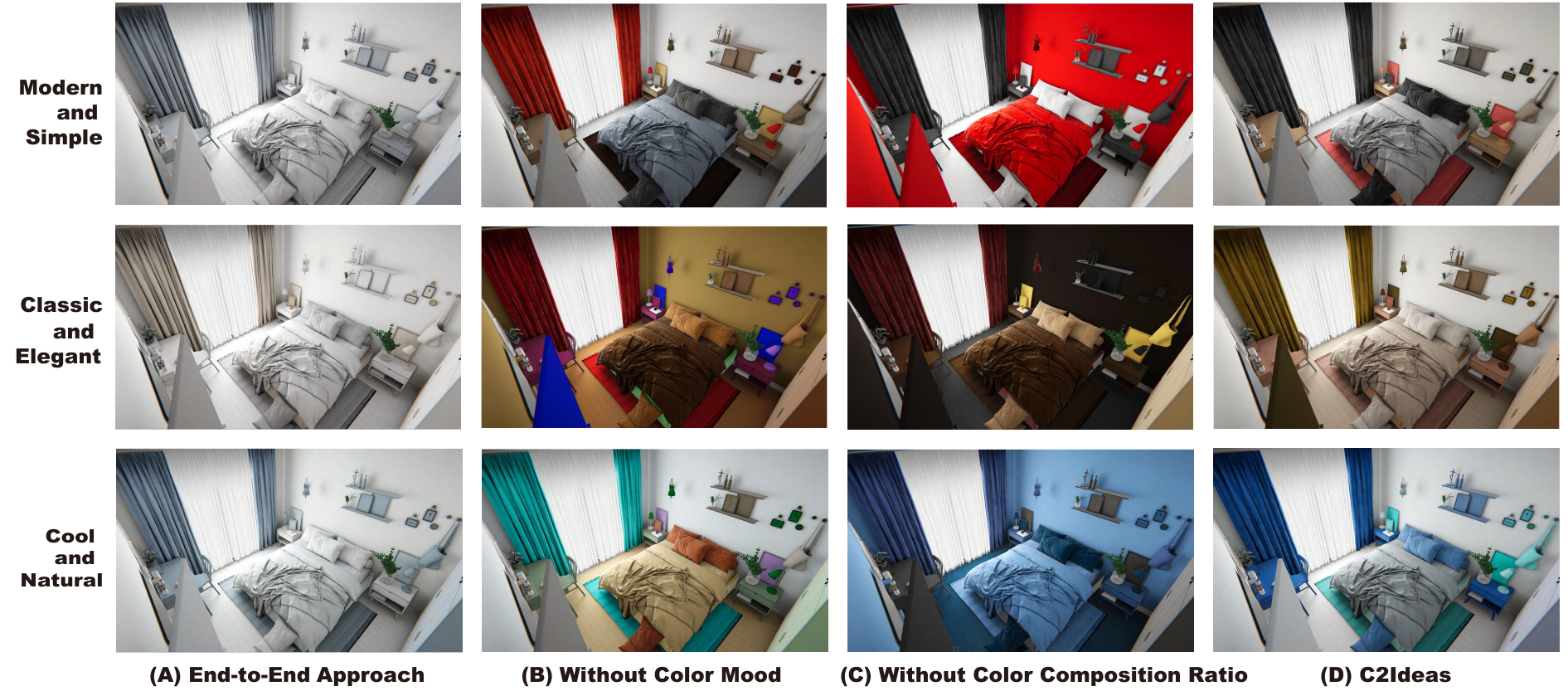}
    \caption{Examples of the result generated using different approaches. (A) End-to-end approach using LLM only, (B) Variation that removes the domain knowledge of color Mood, (C) Variation that removes domain knowledge of color composition ratios, (D) \tool.}
    \vspace{-4mm}
    \label{fig:compare}
\end{figure}

\subsubsection {Experimental Design}
To evaluate the effectiveness of our approach, we conducted a user study to compare the color scheme of the interior generated by \tool with three variants that exclude specific domain knowledge.
The baseline known as "End-to-end" lacks all domain knowledge for interior color design; it does not include the process of the chain of thought, which directly outputs color scheme results based on inputs.
We conducted ablation experiments in which subsequent variants selectively exclude domain knowledge like "color mood" (to ensure reasonable color generation and prevent conceptual drift) and "color composition ratios" (to maintain visual harmony and avoid overly bright or saturated colors).

\textbf{Data.}
\re{
We generate the testing dataset using three scenes and five styles with the four approaches.
We include one simple scene and two complex scenes, which encompass common scenes within the realm of home interior design.
    The study room has 12 basic items such as a desk and chair (Figure \ref{fig:cases} A).
    The living room and bedroom, as complex scenes, include material and texture details and retain the normal map for realism.
    The living room features diverse furniture, including 13 items for color assignment (Figure \ref{fig:cases} B).
    The bedroom is the most complex, with 21 colorizable elements such as beds, cabinets, frames, and decorative items (Figure \ref{fig:cases} C).
}

In order to demonstrate \tool's feasibility of handling diverse user intent, we opted to include a representation that spread across a diverse range of semantics.
\re{
    Thus, we refer to the typical color semantic mapping proposed by Kabayashi~\cite{kobayashi1981aim, lee2006development}, and we choose five styles that span the color sentiment on dimensions of "warm-cool" and "soft-hard".
    We then determined the styles together with the domain experts, to ensure they are practical and representative in the real-world interior design.
}
We finalized five styles as users' initial intent, encompassing \q{Modern and Simple}, \q{Classical and Elegant}, \q{Cool and Natural}, \q{Warm and Cozy} and \q{Energetic and Dynamic}.
The descriptions of these input styles deliberately avoid explicit color references and professional stylistic representations, as we wanted to make them more relevant to real-world scenes where users are often unable to provide direct specifications regarding color or thematic preferences for interior design.
As a result, the descriptions furnished to designers can only convey a general sense or feeling of the intended design.
This motivated our choice of employing higher-level and more abstract descriptions to assess whether \tool can bridge the gap between homeowners and the world of interior design.
\re{Figure \ref{fig:cases} presents the results generated by \tool for the three room settings based on the five styles.}

For each style-environment pairing, three distinct results were generated using each approach, without \re{user modification}.
This methodology was adopted to mitigate potential biases from outliers or individual user preferences, ensuring a more robust evaluation of the approaches.
A subset of our test examples can be found in Figure \ref{fig:compare}.
For a comprehensive view of the entire testing dataset, kindly consult the Supplementary Material.

\textbf{Participants.}
A total of 14 participants including 6 male and 8 female, aged between 22 and 41 (M=26.14, SD=5.15), were recruited from a professional design community.
These participants hailed from diverse design backgrounds, encompassing both academic and professional realms.
Specifically, 4 participants (P1, P3, P5, P7) held university degrees in design, another 4 (P2, P11, P12, P13) possessed professional design work experience, and 6 participants (P4, P6, P8, P9, P10, P14) had both academic qualifications and professional experience in the field.
Notably, all participants demonstrated a foundational understanding of color design and were familiar with the design process.

\textbf{Procedure.}
The study was structured in the sequence of introduction, experiment, and questionnaire.
Initially, participants were given a 15-minute introduction about the motivation behind our work, focusing on the concepts of color mood and color composition ratios.
We also highlighted the four evaluation metrics, as depicted in Figure~\ref{fig:UserStudyResult}, to assist participants in distinguishing between the criteria.
The experiment commenced once participants confirmed their understanding of the tasks and the principles of reasonable interior color design.
During the main task, participants were presented with a set of interior color designs for identical scenes and styles produced by the End-to-end approach, the approach without color mood, the approach without color composition ratios, and \tool.
\re{
    In the experiment, participants were unaware of the specific identities of the four methods, referred to as Design 1 to 4, to prevent bias from preconceived notions. 
    The order of these designs was randomized for each participant to mitigate order effects.
    Each participant experienced the same sequence of designs consistently across different scenes and styles, to facilitate comparative qualitative feedback. 
    The specific sequence varied between participants, balancing comparative analysis with bias minimization due to order consistency. 
}
Participants could freely adjust the scene's angle and viewport on the desktop.
Upon reviewing all results, they were directed to rate them based on the four evaluation metrics through a questionnaire, as illustrated in Figure~\ref{fig:UserStudyResult}.
During the study, the participants are asked to think aloud during the process of filling out the questionnaire, giving their reasons for the evaluation results.
These verbal responses are recorded and subjected to analysis in the feedback section of our research.

\textbf{Evaluation Criteria.}
The questionnaire, inspired by prior work's evaluation metrics~\cite{lin2020c3}, prompted participants to rate their agreement with the evaluation criteria on a 5-point Likert scale, where 1 signifies \q{completely disagree} and 5 denotes \q{completely agree.} The criteria encompassed:
\begin{itemize}
    \item \textit{\re{Outcome Relevance}}: \re{Participants gauged the relevance of the generated results to the input styles, assessing the alignment between predetermined input styles and generated results.
              For example, evaluating whether the resulting interior color scheme fits the style of \q{modern simplicity}}.
    \item \textit{Preference}: This assessed user satisfaction with the results, allowing for subjective evaluations.
    \item \textit{Reasonableness}: The participant evaluated the generated results by considering both color composition ratios as well as color mood dimensions, which represent the extent to which the generated results take into account domain knowledge.
    \item \textit{Diverse and Inspiring}: This indicator assesses the diversity and inspirational potential of the generated results, emphasizing the ability of the approach to provide color schemes that can bring new ideas to the designer.
\end{itemize}

In total, each participant completed 60 trials in each experiment (4 approaches $\times$ 5 styles $\times$ 3 scenes).
On average, each participant spent 1-1.5 hours for the study and received a compensation of \$10 for their participation.

\subsubsection{Result and Analysis}
\begin{figure}
    \centering
    \vspace{-4mm}
    \includegraphics[width=0.98\linewidth]{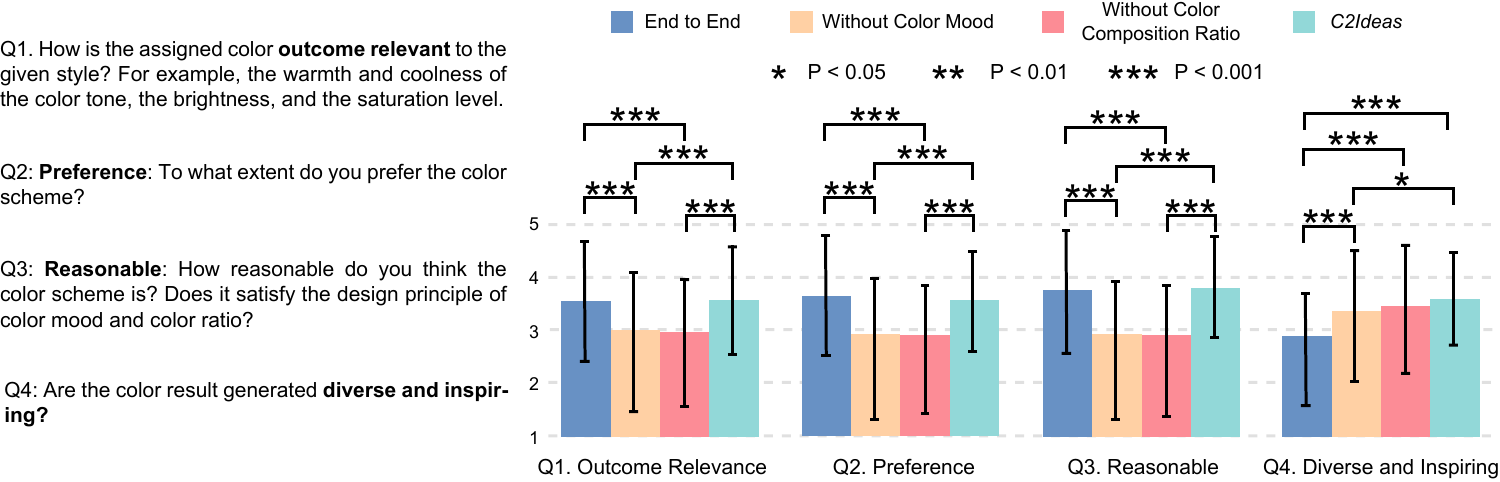}
    \caption{User ratings of the generated results by End-to-end, color mood removal, color composition ratio removal, and \tool.}
    \vspace{-4mm}
    \label{fig:UserStudyResult}
\end{figure}

We collected a total of 3360 answers (60 trials $\times$ 4 questions $\times$ 14 participants).
After the normality test, the nonparametric Friedman's test is used to show the significant effect on the evaluation results of the four approaches, as shown in Figure~\ref{fig:UserStudyResult}.
We also used the Conover test to perform pairwise comparisons between the four approaches.
Significant values are reported for $p <.05(\ast)$, $p <.01(\ast\ast)$, $p <.001(\ast\ast\ast)$.
The findings are summarized as follows:

\textbf{\re{Outcome Relevance} (Q1).}
\tool's \re{outcome relevance} significantly outperforms the two approaches that eliminate domain knowledge ($d.f. = 209, \ p <.001 $), and so does the End-to-end approach ($d.f. = 209, \ p <.001 $).
Although no significant difference was found in \re{outcome relevance} between \tool and the End-to-end approach,
\tool is rated as the best ($mean=3.54$, $SD=0.98$), followed by End-to-end approach ($mean=3.53$, $SD=1.09$).
The two approaches that remove domain knowledge performed the worst.
The approach without color composition ratios got the worst score ($mean=2.76$, $SD=1.16$), while the one with color mood removed scored slightly higher ($mean=2.78$, $SD=1.26$).

\textbf{Preference (Q2).}
Participants rated \tool significantly higher than both color mood removals ($d.f. = 209, \ p < .001 $), so did the End-to-end approach ($d.f. = 209, \ p < .001 $).
Although the End-to-end approach had a slightly higher average (mean=$3.64$, SD=$1.09$), \tool (mean=$3.53$, SD=$0.91$) demonstrated greater stability.
Additionally, color mood removals ($mean=2.65$, $SD=1.30$) only marginally outperformed color composition ratios removals ($mean=2.63$, $SD=1.17$) and did not demonstrate a statistically significant advantage.

\textbf{Reasonableness (Q3).}
For the criterion of reasonableness, \tool received the highest score and demonstrated a significant improvement when compared to both approaches that lacked color mood ($d.f. = 209, \ p < .001 $) or color composition ratios ($d.f. = 209, \ p < .001 $).
In the second position was the End-to-end approach, which also exhibited significantly better results compared to the removal of color composition ratios ($d.f. = 209, \ p < .001 $) and the removal of color mood ($d.f. = 209, \ p < .001 $).
Color composition ratio removals (mean=$2.60$, SD=$1.17$) resulted in unreasonable outcomes and were rated the lowest.
While color mood removals (mean=$2.64$, SD=$1.26$) produced somewhat more reasonable results, this difference was not statistically significant when compared to color composition ratios removals.

\textbf{Diverse and Inspiring (Q4).}
\tool is considered to be the most diverse and capable of inspiring designers, which significantly outperforms the End-to-end approach ($d.f. = 209, \ p <.001 $) and the approach without color mood removal ($d.f. = 209, \ p <.05 $).
The End-to-end approach has no advantage in terms of diversity and inspiring power and significantly lags behind the other three approaches ($d.f. = 209, \ p <.001 $).
The domain knowledge removals for the first time are rated higher than the End-to-end approach in this whole user study.
Compared to color composition ratios removals (mean=$3.39$, SD=$1.16$), \tool (mean=$3.56$, SD=$0.86$) are rated higher and more stable in terms of diversity and inspiration.

\subsubsection{Feedback}
In our qualitative analysis of participants' feedback, we utilized open coding to identify recurring styles from participants' expressions.
Two researchers reviewed and categorized participants' comments into thematic clusters, iteratively refining the coding scheme for data saturation.
The findings are summarized as follows:

\textbf{End-to-end: Stable but Monotonous.}
In evaluating End-to-end's \re{relevance with the given style}, participants primarily rated it highest within the \q{Modern and Simple} style.
They appreciated \q{the overall color coordination and simplicity}, which aligns with this style (P11).
However, in other styles, End-to-end results were not as dominant and were sometimes perceived as \q{overly simplistic}, deviating from the intended atmosphere (P7).
From a personal preference perspective, participants with design backgrounds favored the overall simplicity (P1) and believed it would \q{appeal to ordinary users} as well (N=4).
They praised End-to-end reasonableness for adhering to color composition ratio restrictions and avoidance of overly saturated or bright colors for reasons (N=5).
However, when it came to diversity and inspiration, participants felt that End-to-end results were limited and lacked variety (N=7).
They described it \q{tries to avoid mistakes in a very safe range} (N=5), like a \q{lazy designer} (P5), who only gives a design scheme that is universal for various styles and lacks creativity.
P4 even perceived End-to-end as \q{constrained within a fixed range of color and saturation choices that hindering creative freedom}.

\textbf{Domain Knowledge Removals: Bold but Chaotic.}
These knowledge removal approaches, while bold in creativity, tend to result in overall chaos.
There was a clear two-level differentiation in personal preference evaluations of the results of these two approaches, with some participants with niche preferences (N=5) appreciating the novel designs whilst others (N=6) strongly disliked them.
Across all scenes and styles, \re{participants consistently assigned low reasonableness ratings to both approaches due to their abrupt color matching (N=3) and inconsiderate color composition ratios (N=7).}
While some participants (N=6) attributed the unreasonableness to \q{overly exciting color selections}, others (N=3) believed it stemmed from \q{chaotic color choices}.
In terms of diversity and inspiration, these approaches introduced uncommon color combinations that sparked new ideas (N=2), whilst this inspiration often proved challenging to apply in real-world scenes, serving as a cautionary \q{don't do this} type of inspiration (P5).
Both methods show an advantage in all evaluation perspectives in the style of \q{Energetic and Dynamic}, where bold color choices are implied in the style itself (N=4).
P12 described them as \q{designers who are trying hard}, with one losing color mood guidance, causing directional errors in color selection, and the other ignoring color composition ratio constraints, resulting in color overuse.

\textbf{\tool: Balanced and Promising.}
\tool was found to have the \q{best overall color scheme} (N=8).
In comparison to the other two domain knowledge-free approaches, it was noted for \q{using lively colors while maintaining overall control} (P2), and in contrast to End-to-end, it \q{stays within a stable range while attempting diversity} (P3).
Several participants perceived \tool as a balance between End-to-end and the knowledge removal approach (N=4), combining the strengths of both.
Our results effectively balance creativity and reasonableness, making them more practical for real-world use (N=2).
As P5 pointed out, \q{as someone needing to renovate, I'm more inclined to use the results generated by your approach,} while P1 remarked, \q{as a designer, your designs are useful.}
Nonetheless, refinements may be required for \tool's results to achieve a directly usable color scheme, given that the misuse of a single color can disrupt the overall atmosphere.
As P6 observed, \q{while it's got a pretty good mix of colors, they're too close in saturation, lightness, and darkness to stand out}.
Furthermore, P12 noted that this issue might arise because \q{the use of complementary and accent colors in the interior design itself poses a significant challenge to a designer's skills, and it becomes even more complex for a machine to comprehend how they interact}.

\subsection{Experts Feedback on \tool Interface}
In this study, we conducted expert interviews to assess the \re{usability} of \tool.
Participants were instructed to utilize \tool for generating color schemes and evaluate their \re{controllability, intent alignment, and general usability} through a structured questionnaire, which was complemented by semi-structured interviews.

\begin{table}[h]\tiny
    \caption{Demographic Information of Expert Interview Participants}
    \resizebox{\linewidth}{!}{
        \begin{tabular}{cccccccccccc}
            \toprule
            \textbf{Alias} & \textbf{Age} & \textbf{Gender} & \textbf{Occupation}                & \textbf{Affiliation}     & \textbf{Experience (Years)} \\
            \midrule
            E1             & 28           & Male            & PhD Student (Art)                  & Comprehensive University & 3-4                         \\
            E2             & 30           & Female          & Soft Furnishings Interior Designer & Interior Design Company  & 9-10                        \\
            E3             & 41           & Male            & Soft Furnishings Interior Designer & Interior Design Company  & 17-18                       \\
            E4             & 39           & Female          & Design Programme Professor         & Art College              & 21-22                       \\
            \bottomrule
        \end{tabular}
    }
    \label{tab:expert}
\end{table}

\subsubsection{Study Design}
We enlisted four participants for the expert interview phase. In our pursuit of obtaining comprehensive professional feedback, we recruited interior designers through the online designer platform.
All participants possessed ample design experience, and their demographic information can be referenced in Table~\ref{tab:expert}.

At the outset of the study, we acquainted each participant with the functions and associated interactions of \tool.
Once they were familiar with the system, we presented them with two tasks:
(1) Task 1 required participants to select one of the five design styles and generate a color scheme for an interior space. They were permitted to modify the color scheme iteratively until satisfied with the outcome;
(2) Task 2 encouraged participants to freely explore \tool, inputting their personalized intent and trying more possibilities.
The session concluded with a semi-structured interview, gathering their feedback and recommendations for \tool.
Each study lasted between 1 to 1.5 hours. Participants received a compensation of \$15 for their time and insights.
Their responses were recorded for subsequent analysis.

\subsubsection{Result and Analysis}
Our semi-structured interview with the experts highlighted their experiences and insights on using \tool.

\textbf{Alignment with Existing Design Workflow.}
The stepwise generative process of the system aligns well with conventional interior design methods.
Most experts (N=3) found this workflow, which conceptual association, color scheme selection, and final design, to be compatible with their usual practices.
All four experts unanimously agreed that the stepwise generation significantly enhances control within the design process.
Expert E3, however, occasionally diverged from this approach by having a pre-visualized design in mind, which allowed them to skip some generative steps.
As stated by Expert E3 \q{had already placed furniture with color in the space in mind when conceptualizing the overall room layout.}
Despite this divergence, Expert E3 found our tool to be valuable during creative lulls.
While experts did not consider direct model generation using an end-to-end approach to fulfill the requirements of a design tool.
\re{Expert E2} noted that End-to-end generation might cater to certain customer preferences due to its ability to \q{generate a large number of results quickly, and not much modification is required to reach the borderline}.

\re{
    \textbf{Alignment with User Intent.}
    The effectiveness of \tool in aligning with user intent was affirmed by all experts (N=4).
    The feedback highlighted the system's ability to produce results that closely reflected the users' design objectives, necessitating minimal adjustments.
    Expert E3, upon finding a satisfactory color palette in the third step, remarked, \q{It's possible to find one that's closest to how I feel, with little or no tweaking.}
    Echoing this sentiment, Expert E1 noted, \q{It generates a color that looks like the one I'm thinking of, except there's too much red, just tweak it a bit.}
    A particularly notable response came from Expert E2, who was surprised and impressed by the system's capacity to accurately interpret and replicate the \q{Mondrian Style}, effectively generating the style's iconic red and blue hues.
}

\textbf{Flexible and Comprehensive Interaction.}
\tool accommodates both vague and specific design intents through two interaction modes.
This flexibility in interaction is acknowledged by all four experts, highlighting \tool's support for multiple dimensions and levels of interaction.
Experts noted that modifying the three dimensions of color mood in Idea Prompting was effective in altering the overall color rendering (N=3).
Additionally, the availability of individual color adjustments in the last two steps met the requirement for precise adjustments, a consensus among all four experts.
For well-defined intent, experts lean towards using the GUI. For instance, in the free exploration phase, E2 selected "Mondrian Bedroom" and rapidly achieved her desired result by individually adjusting decorations to the color red. However, the NLI system shines when users offer vague or unclear design feedback, as exemplified by E1's request to "make the interior more colorful."
While the system's natural language interaction (NLI) presents is promising for system adjustments, its current limitations in precisely interpreting text input occasionally result in discrepancies between users' desired and actual changes (E3).
After E3 entered \q{Large area color is too heavy}, the model was modified only to change the floor color to a brighter grey, but not the walls, which are a much larger area.

\textbf{Support both Novices and Experts.}
The system caters to a spectrum of users, from novices to experts.
Novices, when unfamiliar with certain styles or themes, could rely on \tool to curate examples, enhancing their learning curve.
Expert E4 mentioned that when her students first encounter a new style or theme, \q{the fastest way to learn is to find an example to learn the color selection and extract the color composition ratios, which is then learned by imitation.}
For seasoned designers, the AI's color suggestions can help them to see some new color choices and matching possibilities even when they are very familiar with or have already designed a room with the same style (N=3).

\section{Discussion}
\label{sec:discussion}

\subsection{Designer's Expectation for AIGC in Creative Design}
\label{ssec:design}
Our investigation into designers' expectations regarding AI-supported tools yielded intriguing insights.
Contrary to our initial assumptions, the LLM demonstrated an impressive capability to generate reasonable colored scenes.
These results resonated positively with general users, as some participants (N=3) noted that the End-to-end approach could \q{generate designs that can be presented directly to clients without any reservations and will not get yelled at, especially during busy schedules.}
However, for professional designers, merely reasonable and aesthetic appeal falls short of their expectations.
Their aspirations for AI tools extend beyond rapid result generation and automating mundane tasks.
While these designs may be appealing to a non-professional audience, designers often find them \q{conservative and perfunctory}, even questioning if they meet the true essence of `design'.
Designers yearn for AI tools that can be involved throughout the design process, not just as passive assistants but as active collaborators.
They seek inspiration from these tools, hoping for AI's potential to spark diverse and innovative designs, pushing the boundaries of conventional design thinking.
\re{
Such insights can enhance the development of AI-assisted tools, aligning them more closely with human-centric design principles, and fostering a more harmonious relationship between designers and AI~\cite{subramonyam2021towards,yildirim2023creating,feng2023ux}.
}

Our approach and system excel in balancing the generation of reasonable results while providing inspiration for designers.
It offers \q{something out of the ordinary within the realm of reason} (P4), highlighting the importance of this balance for the practical integration of AIGC in design practice.

\subsection{Designer's Contribution to AIGC in Creative Design}
\label{ssec:authoring_tool}

Traditionally, the development of an authoring tool for automated Interior Coloring has been a laborious and formidable undertaking.
This task typically requires the substantial construction of an extensive database comprising myriad color combinations or the formulation of intricate rules grounded in color theory.
Despite these efforts, the resultant outcomes frequently fall into the categories of either \q{mediocre} or \q{chaotic}, rendering them arduous to apply effectively within the realm of real-world design practice.
The introduction of the LLM has marked a transformative milestone in this landscape, as it can rapidly generate colored scenes by replacing the need for extensive datasets with its pre-trained model and complex rule sets with natural language prompts.

Nonetheless, it is imperative to recognize that the LLM is not a universal panacea for all design challenges.
Designers play a pivotal role in integrating AIGC into the creative design process.
In this work, we have forged close collaborations with designers as domain experts.
We have meticulously delved into design rationale and processes, engaging in numerous discussions to ensure that our system aligns closely with designers' real-world design practices.
Following a user-centric design approach, we have painstakingly designed, evaluated, and refined our framework and system to make it seamlessly integrate with designers' workflows.
With the invaluable contributions of designers, we can harness the potential of the LLM to enhance and support the creative design process, rather than generating \q{standard} results based on the general model.

\subsection{Limitations and Future Work}

\textbf{Enhancing Scene Complexity and Adaptability.}
Our current work has primarily focused on scenarios of varying complexity while assuming fixed layouts in the design process.
We only conducted tests on a limited set of three scenarios, spanning a spectrum from simplicity to moderate complexity.
However, real-world design challenges often involve dynamic changes in furniture arrangements and multifaceted scenarios that require intricate solutions.
In future research, we aim to expand our approach to address both the adaptability of scenes, accommodating dynamic layout adjustments, and the intricacies of more complex design scenarios.
This holistic approach will better align with the multifaceted nature of real-world design practice, offering designers a more flexible and comprehensive toolset for creative design.

\textbf{Incorporating Subtle Design Rationales and Cultural Influence.}
Design, being a complex field, involves a vast array of design rationale and knowledge.
\re{
    While we have integrated two primary aspects of color theory—color ratio and color mood—into our interior color design process, practical design often involves richer and more personalized guidelines.
    For instance, E4 noted that in terms of color composition ratio, the 6:3:1 ratio can be adjusted to some extent based on the specific colors chosen.
    Future developments should consider encompassing a broader range of color composition ratio paradigms or enabling designers to customize.
    This flexibility would empower designers to exercise more creative control over their projects, fostering personalized and innovative interior designs.
    Additionally, experts expressed the need for more comprehensive and specialized domain knowledge (N=3).
    For instance, E1 suggested we expand the existing knowledge base to include aspects like "color matching" and "style integration".
    The cultural influence on design is another aspect that we have not yet considered.
    In some cultures, colors hold unique symbolic meanings, for example, \q{black} represents horror in Western culture while signifying peace and mystery in some Eastern cultures.
    Our future work will delve deeper into these subtler aspects of design rationale, aiming to provide a more comprehensive and culturally aware framework for creative design.}


\textbf{Balancing Diversity and Reasonability.}
Within the creative design process, the imperative task of reconciling diversity and reasonability holds a central position.
Designers actively seek inspiration from a diverse array of generated results, while simultaneously demanding that these outcomes retain a level of practicality and relevance within the context of real-world design scenarios.
This is what designers mean when they refer to \q{inspiring design}, and it forms a critical cornerstone of our research focus.
In our forthcoming investigations, we will explore advanced methodologies to effectively navigate this delicate equilibrium.
These strategies may encompass providing designers with explicit control over result diversity or harnessing a designer's prior work as a point of reference, facilitating the creation of highly personalized and contextually sensitive design solutions.

\re{
    \textbf{Support Image Modality.}
    In our current work, we concentrate on addressing users' vague intentions that are typically expressed in text as the primary input modality.
    Nevertheless, we recognize that designers often use visual references, like mood boards or images from previous projects, to communicate their design intent.
    Moving forward, we plan to expand our framework to incorporate these visual references, allowing designers to utilize their existing design assets for creating more personalized and context-relevant design solutions.
}
\label{ssec:future_work}

\section{Conclusion}
We have presented \tool, a pioneering system that facilitates the creative ideation of color schemes in interior design.
This system leverages an intent-aligned and domain-oriented large language model to bridge the gap between automated design and user intention, addressing a known misalignment in current research.
\tool facilitates the creative ideation process through a three-stage approach: Idea Prompting, Word-Color Association, and Interior Coloring, which align well with the existing design workflow.
Augmented by its user-centric interactive interface, \tool grants users not only the capability to fine-tune their designs but also clarity in design reasoning. 
Through a comprehensive set of indoor cases and user studies, \tool has demonstrated its capability and effectiveness in interior color design. 
Designers affirmed our approach's balance between reasoning and creativity, and have notably acknowledged and appreciated the system's interactive functionalities.
We hope that this work will provide insights into how to use LLM for enhancing creative color design and fostering cross-disciplinary collaboration with designers.

\bibliographystyle{ACM-Reference-Format}
\bibliography{sample-base}

\end{document}